\documentclass[aps,twocolumn,superscriptaddress,nofootinbib,prb]{revtex4-2}

\usepackage{graphicx}
\usepackage{placeins} 
\usepackage{amssymb,amsmath,amsfonts,bm}
\usepackage{subcaption}
\usepackage[normalem]{ulem}
\usepackage{hyperref}
\usepackage[dvipsnames]{xcolor}
\hypersetup{urlcolor=RoyalBlue, colorlinks=true,citecolor=RoyalBlue,linkcolor=RoyalBlue} 
\usepackage{physics}
\usepackage[final]{showkeys}
\usepackage{float}

\begin{document}
\title{Exact Rokhsar–Kivelson points hosting first-order phase transitions in $S=1$ compass models}

\author{Jay Pandey}
\email{jay.pandey@tifr.res.in}
\affiliation{Department of Theoretical Physics,
	Tata Institute of Fundamental Research,
	Mumbai 400005, India}
\date{\today}
\begin{abstract}
We study $S=1$ compass models on the square and kagome lattices with bond-dependent bilinear and biquadratic interactions parametrized by an angle $\theta$. For compatible geometries, we determine the exact ground-state manifolds throughout the interval $\pi/4\leq\theta\leq3\pi/4$. At each endpoint, a product of on-site unitaries maps the Hamiltonian to the Laplacian of a configuration graph $\mathcal{G}$ in the $S=1$ Cartesian basis. These endpoints are exact Rokhsar–Kivelson (RK) points, with one ground state per dynamically connected component of $\mathcal{G}$. Each such state is an equal-amplitude superposition in the transformed basis and acquires configuration-dependent phases upon transforming back to the original basis. Both models possess $m=O(L)$ independent and mutually commuting conserved $\mathbb{Z}_2$ line charges, where $L$ is the linear system size, forcing a ground-state degeneracy of at least $2^m$ at the RK endpoints. Throughout the interior region, $\pi/4<\theta<3\pi/4$, the ground-state manifold on each lattice is instead spanned by two product states, both annihilated by the Hamiltonian. On both lattices, each RK endpoint also hosts a first-order transition into this intervening interior phase spanned by the two product states with long-range order.
\end{abstract}

\maketitle
\section{Introduction}
Exactly solvable models are often helpful in understanding otherwise mysterious properties of interacting many-body systems. Familiar examples include the AKLT chain~\cite{AKLT1987},
the Kitaev honeycomb model~\cite{Kitaev2006}, and quantum
dimer models (QDMs) at their Rokhsar--Kivelson (RK)
points~\cite{RokhsarKivelson1988,MoessnerSondhi2001,Misguich2002,Syljuasen2006,MoessnerRaman2011}. The celebrated Kitaev model~\cite{Kitaev2006} is but one example of a more general class of models called compass models, in which the coupling on a bond depends on its orientation and selects a different spin component along each lattice direction~\cite{Kugel_1982,NussinovVandenBrink2015,Nussinov_Fradkin_PRB_2005,Dorier_etal_PRB_2005}. In several compass models, this structure supports subsystem
symmetries that constrain ordering and dynamics~\cite{Xu_Moore_2004,Dorier_etal_PRB_2005,Orus_etal_2009,Morita_etal_PRB_2018,Yang_etal_PRB_2020,Kimchi_Vishwanath_2014,Li_etal_2024}. This can partition the Hilbert space into many dynamically disconnected sectors, with further disconnected parts possible within a sector~\cite{Sala_2020,Khemani_2020,Moudgalya_2022,Budde_preprint_2026}. Such conservation laws tied to the lattice geometry, can impose
mobility constraints reminiscent of fracton systems~\cite{Vijay_2016,Nandkishore_Hermele_2019}.

The RK points alluded to earlier were originally introduced in QDMs, where the ground states are equal-amplitude superpositions of dimer coverings connected by local resonance moves~\cite{RokhsarKivelson1988,MoessnerSondhi2001}. Later it was realized that one can construct similar RK points in other systems like quantum vertex and coloring models, establishing a broader connection between exact quantum ground states and classical statistical mechanics~\cite{Henley_2004,Castelnovo_2005}. At an exact RK point the negative of the Hamiltonian  acts as the  generator of a continous-time classical Markov process on each connected component of a graph. Consequently, each component hosts an equal-amplitude ground state, and its excitation energies coincide, up to an overall rate scale, with the relaxation rates of the corresponding classical process~\cite{Henley_2004,Castelnovo_2005}.
 
 Our work has its origins in Ref.~\cite{Raja_Ganesh_2026} which constructed exponentially many exact ground states of a bilinear-biquadratic $S=1$ Kitaev chain using spin fractionalization. This work of  Ref.~\cite{Raja_Ganesh_2026} raises the natural question of whether analogous exactly solvable points exist in higher-dimensional $S=1$ models.  We address this question here, and in doing so, bring together RK physics and subsystem-symmetry induced Hilbert-space decomposition in $S=1$ compass models on the square and kagome lattices as concrete examples. The kagome compass model is known more popularly as the kagome Kitaev model. On top of the standard bond-dependent bilinear interactions in such models we also include biquadratic couplings of the same spin component on each bond, with their relative strengths parametrized by an angle $\theta$. Our approach avoids spin fractionalization and, subject to geometric compatibility conditions specified later, applies to $S=1$ models without restricting the lattice coordination number. It not only yields exact ground-state wavefunctions but also makes a statement about the entire many-body spectrum of these models by showing the presence of exact RK points. It also applies to the $S=1$ Kitaev chain.
 
In quantum magnets, RK constructions include models with effective ring-exchange dynamics in constrained low-energy manifolds~\cite{BalentsFisherGirvin2002,HermeleFisherBalents2004}, as well as spin Hamiltonians designed to support exact RVB or vertex-model ground states~\cite{CanoFendley2010,BalasubramanianGalitskiVishwanath2022}. These approaches typically involve interactions acting on more than two sites or rely on a constrained low-energy description. Here we establish a direct route to RK physics in two-dimensional $S=1$ compass models having only nearest-neighbour two-site bilinear and biquadratic interactions.
On compatible geometries, a product of sublattice-dependent on-site unitaries brings the full Hamiltonian into an adjacency-degree representation on a configuration graph in the Cartesian basis. This representation holds for arbitrary $\theta$ and reveals exact RK points at $\theta=\pi/4$ and $3\pi/4$.
The construction requires neither a low-energy projection nor additional multispin interactions and directly relates the ground-state manifolds to configuration-space connectivity.

In these compass models on compatible geometries the ground-state manifold is known exactly throughout $\pi/4\leq\theta\leq3\pi/4$. In the interior of this interval, both models have two product states in the ground state manifold. Each is annihilated by every term in the Hamiltonian, whose positive semidefiniteness in this interval guarantees that these states minimize the energy. At both endpoints ($\theta=\pi/4$ and $\theta=3\pi/4$) on the two lattices, a product of on-site unitary operators maps the Hamiltonian to the Laplacian of a configuration-graph $\mathcal{G}$~\cite{Mohar1991,GodsilRoyleLaplacian2001} whose vertices represent  product configurations in the $S=1$ Cartesian basis. Each connected component of $\mathcal{G}$ supports one state in the ground-state manifold, an equal-amplitude superposition over the configurations in the connected component in the transformed basis, with configuration-dependent phases in the original basis~\cite{GodsilRoyleLaplacian2001,Mohar1991}.

 Both the kagome and square $S=1$ compass models have $m=O(L)$ independent and mutually commuting conserved $Z_2$ line charges and hence $2^m$ sectors~\cite{NussinovVandenBrink2015,Dorier_etal_PRB_2005,Yang_etal_PRB_2020}. Thus, the subsystem symmetries enforce at least $2^{m}$ connected components and hence a corresponding ground-state degeneracy exponential in the linear system size at the RK points. These ground state wavefunctions have a natural PEPS representation~\cite{VerstraeteCirac2004PEPS,Verstraete_etal_PRL_2006,
	VerstraeteMurgCirac2008,Cirac2021MPSPEPS} of bond dimension two on both the lattices.
These endpoints are therefore exact RK points, analogous to those in QDMs, with the exponentially many sectors labelled by the line-charge values. The RK endpoints also host first-order transitions bounding a phase with long-range order on both the lattices.

\section{Model}
\begin{figure*}
	\centering
	\includegraphics[width=\linewidth]{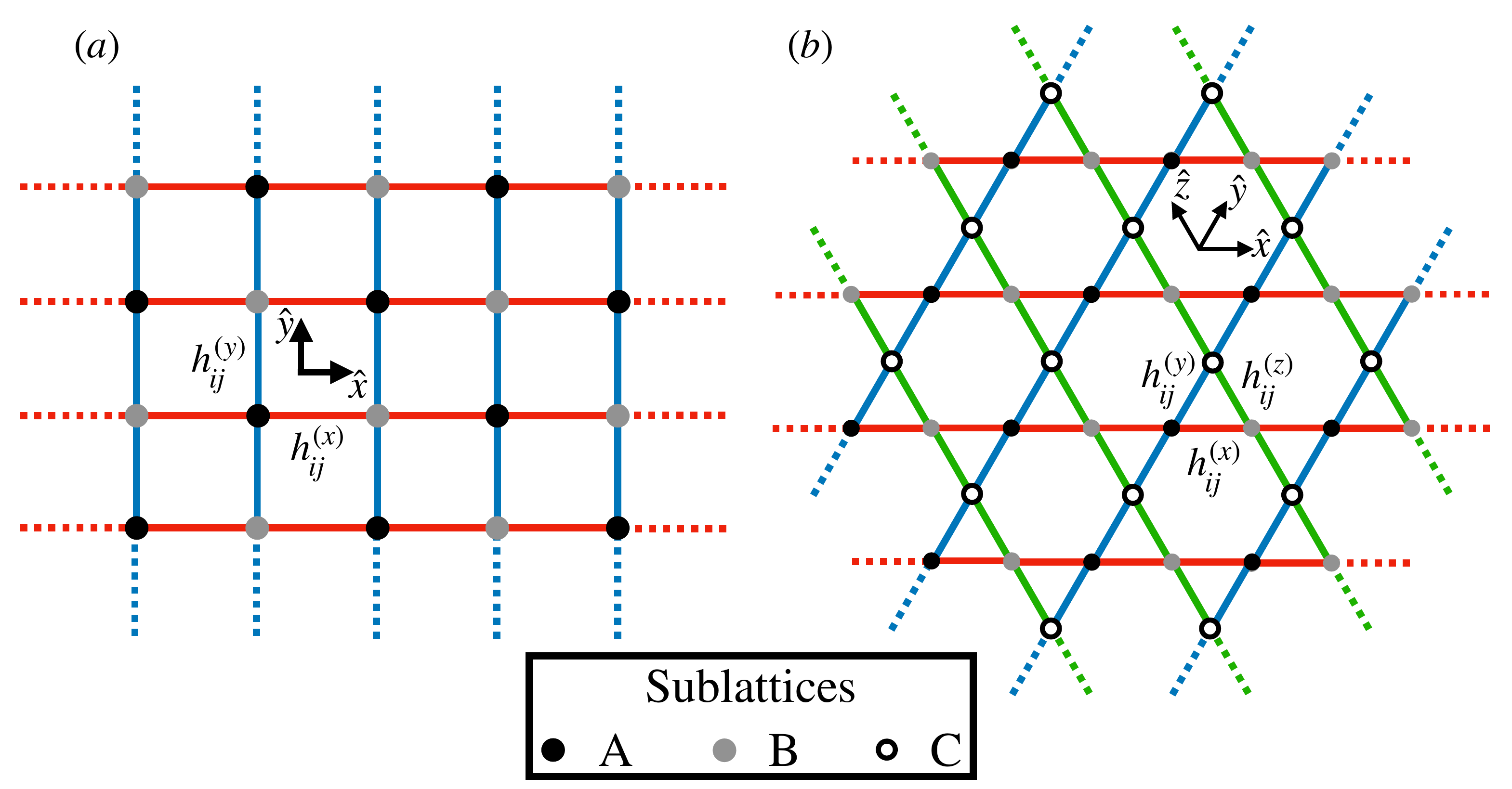}
	\caption{$S=1$ compass models on the square (left) and kagome (right) lattices. Filled black, filled gray, and open black circles denote the $A$, $B$, and $C$ sublattices, respectively. The square lattice has a single-site unit cell, whereas the kagome lattice has a three-site unit cell forming an upward-pointing triangle. Red, blue, and green bonds denote $x$, $y$, and $z$ interactions, respectively, with couplings on the bonds
		${h}^{\alpha}_{ij}(\theta)
		= \cos\theta\,\hat{S}^{\alpha}_{i}\hat{S}^{\alpha}_{j}
		+ \sin\theta\,(\hat{S}^{\alpha}_{i}\hat{S}^{\alpha}_{j})^2$.
		The square lattice contains only $x$ and $y$ type bonds.}
	\label{fig:CompassKitaevSchematic}
\end{figure*}

Each lattice site hosts an $S=1$ spin, and we consider both open and periodic boundary conditions (OBC and PBC). The square lattice has a single-site unit cell, while the kagome lattice has a three-site unit cell defined as sites of an upward-pointing triangle. On the square lattice, we allow arbitrary integer dimensions $L_x,L_y\geq2$ for OBC and restrict both $L_x$ and $L_y$ to even values for PBC. On the kagome lattice, we allow arbitrary positive integer values of $L_x$ and $L_y$, both greater than $1$, for both OBC and PBC ($L_x$ and $L_y$ are taken along the two primitive directions of the kagome lattice). Each bond $b$ of the two lattices is assigned an orientation-dependent label $\alpha_b\in\{x,y,z\}$ and it couples the corresponding component of the $S=1$ spins located at its endpoints $i_b$ and $j_b$ through bilinear and biquadratic interactions:
\begin{equation}
	\hat{H}(\theta)
	= \cos\theta \sum_b \hat{S}^{\alpha_b}_{i_b}\hat{S}^{\alpha_b}_{j_b}
	+ \sin\theta \sum_b
	\bigl(\hat{S}^{\alpha_b}_{i_b}\hat{S}^{\alpha_b}_{j_b}\bigr)^2 ,
	\label{eq:CompassHamiltonian}
\end{equation}
where the sum runs over all bonds $b$ of the lattice and $\theta$ parametrizes the relative strengths of the bilinear and biquadratic terms. The square lattice contains only $x$- and $y$-type bonds, whereas the kagome lattice contains all three bond types, as illustrated in Fig.~\ref{fig:CompassKitaevSchematic}.

Two transformations relate different regions of the parameter space in Eq.~\eqref{eq:CompassHamiltonian}. First, $\hat{H}(\theta+\pi)=-\hat{H}(\theta)$, so a shift by $\pi$ reverses the spectrum. Second, a unitary $\hat{V}$ composed of on-site $\pi$ rotations reverses the sign of every bilinear term while leaving the biquadratic terms invariant, giving $\hat{V}^{\dagger}\hat{H}(\theta)\hat{V}=\hat{H}(\pi-\theta)$. For the bond assignments in Fig.~\ref{fig:CompassKitaevSchematic}, $\hat{V}=\prod_{i\in A}e^{i\pi\hat{S}^{z}_{i}}$ on the square lattice and $\hat{V}=\prod_{i\in A}e^{i\pi\hat{S}^{z}_{i}}\prod_{j\in B}e^{i\pi\hat{S}^{x}_{j}}$ on the kagome lattice, where $A$ and $B$ denote the sublattices. The full spectrum can therefore be reconstructed from $\theta\in[0,\pi/2]$. Within $\theta\in[0,\pi]$, the ground-state phase diagram is symmetric about $\theta=\pi/2$; in particular, the points $\theta=\pi/4$ and $3\pi/4$ are unitarily equivalent.

\section{Cartesian basis and bond kinematics}
We work throughout in the $S=1$ time-reversal-invariant Cartesian basis $\{\lvert x\rangle,\lvert y\rangle,\lvert z\rangle\}$~\cite{Sen_etal_PRB_2010,SharmaPandeyInPreparation}, in which
\begin{equation}
	\hat{S}^{\alpha}\lvert \beta\rangle
	= i\epsilon_{\alpha\beta\gamma}\lvert \gamma\rangle,
	\qquad
	e^{i\pi\hat{S}^{\alpha}}\lvert \beta\rangle
	= (2\delta_{\alpha\beta}-1)\lvert \beta\rangle.
	\label{eq:DirectorRules}
\end{equation}
In this basis, each bilinear bond operator maps a Cartesian product configuration of the spins at its endpoints to a single other product configuration or annihilates them, while each biquadratic bond operator is diagonal. Since $\hat{S}^{\alpha}\lvert\alpha\rangle=0$, a bond of type $\alpha$ annihilates any configuration with $\lvert\alpha\rangle$ at either endpoint. We call such a bond \textit{blocked}. The bond acts nontrivially only in the four-dimensional subspace spanned by configurations with neither endpoint in $\lvert\alpha\rangle$ (we call this the \textit{active} subspace). For a cyclic permutation $(\alpha,\beta,\gamma)$ of $(x,y,z)$, this active subspace is $\{\lvert\beta\beta\rangle,\lvert\beta\gamma\rangle,\lvert\gamma\beta\rangle,\lvert\gamma\gamma\rangle\}$. Using Eq.~\eqref{eq:DirectorRules}, the bilinear terms of $\hat{H}$ generate two types of moves within this subspace:
\begin{equation}
	\begin{array}{ll}
		\mathrm{exchange:}
		& \lvert\beta\gamma\rangle \leftrightarrow \lvert\gamma\beta\rangle,
		\quad \mathrm{amplitude}\ +\cos\theta, \\[2pt]
		\mathrm{pairing:}
		& \lvert\beta\beta\rangle \leftrightarrow \lvert\gamma\gamma\rangle,
		\quad \mathrm{amplitude}\ -\cos\theta.
	\end{array}
	\label{eq:moves}
\end{equation}
The biquadratic interaction is proportional to the identity on the active subspace, contributing a diagonal energy $\sin\theta$ to each configuration for every active bond present in it.

Two consequences will be used in what follows. First, the single-bond eigenvalues are $\sin\theta\pm\cos\theta=\sqrt{2}\sin(\theta\pm\pi/4)$, each twofold degenerate, in the active subspace, and zero in the blocked subspace. Second, each active bond permits exactly one move: exchange for $\lvert\beta\gamma\rangle$ or $\lvert\gamma\beta\rangle$, and pairing for $\lvert\beta\beta\rangle$ or $\lvert\gamma\gamma\rangle$ in the Cartesian basis. This construction defines a weighted configuration graph $\mathcal{G}$ with $3^N$ vertices, one vertex for each product state in the Cartesian basis, $|s\rangle=\bigotimes_{i=1}^N|s_i\rangle$, where $|s_i\rangle\in\{|x_i\rangle,|y_i\rangle,|z_i\rangle\}$. The edges carry weights $\pm\cos(\theta)$, and each vertex has a weight $\sin(\theta)D(s)$ due to the biquadratic term, where $D(s)$ denotes the number of active bonds in $|s\rangle$.

\section{Line charge induced Hilbert space decomposition}
The Hamiltonian in Eq.~\eqref{eq:CompassHamiltonian} admits a family of conserved $\mathbb{Z}_2$ charges supported on straight lines, whose number scales with the linear system size~\cite{NussinovVandenBrink2015,Dorier_etal_PRB_2005,Yang_etal_PRB_2020}. Let $\ell$ be a maximal straight chain of bonds of a single type, and let $\tau_i$ denote the common type of the bonds incident on site $i\in\ell$ that lie outside $\ell$. The corresponding line charge is
\begin{equation}
	\hat{Q}_{\ell}
	= \prod_{i\in\ell}e^{i\pi\hat{S}^{\tau_i}_{i}},
	\qquad
	[\hat{Q}_{\ell},\hat{H}(\theta)]=0
	\quad \forall\,\theta.
	\label{eq:Q}
\end{equation}
Each rotation leaves the spin component on bonds outside $\ell$ unchanged, while reversing the component on bonds along $\ell$. Since both endpoints of every bond along $\ell$ are rotated, all bilinear and biquadratic interactions remain invariant.

On the square lattice, $\tau_i$ is constant along each line, yielding $\hat{Q}_{c}=\prod_{j}e^{i\pi\hat{S}^{x}_{(c,j)}}$ for columns and $\hat{Q}_{r}=\prod_{i}e^{i\pi\hat{S}^{y}_{(r,i)}}$ for rows~\cite{NussinovVandenBrink2015,Dorier_etal_PRB_2005}. On the kagome lattice, $\tau_i$ alternates between the two sublattices along each line: an $x$-line ($A$ and $B$ sublattices alternate along it) carries rotations $e^{i\pi\hat{S}^{y}}$ on $A$ and $e^{i\pi\hat{S}^{z}}$ on $B$ sublattices, with the $y$- and $z$-line charges obtained by cyclic permutation~\cite{Yang_etal_PRB_2020}. The line charges satisfy $\hat{Q}_{\ell}^{2}=1$ and are diagonal in the Cartesian basis. Their eigenvalues $q_{\ell}=\pm1$ therefore partition the Hilbert space into dynamically disconnected sectors. An $L_x\times L_y$ square lattice has $L_x+L_y$ independent line charges and hence $2^{L_x+L_y}$ sectors. On the kagome lattice, the three line families provide $L_x+L_y+\gcd(L_x,L_y)$ independent charges under PBC and $2L_x+2L_y-1$ under OBC, yielding $2^{L_x+L_y+\gcd(L_x,L_y)}$ and $2^{2L_x+2L_y-1}$ sectors, respectively. The number of sectors is thus exponential in the linear system size on both the lattices.

On both the lattices the conserved line charges divide the configuration graph $\mathcal{G}$ into disconnected parts labelled by the line charge eigenvalues. Its vertices are the $3^N$ product configurations in the Cartesian basis, and its edges are the allowed moves. Each connected component lies within a line-charge sector, but any given sector can host more than a single such component. The line charges therefore give a lower bound on the number of connected components present in $\mathcal{G}$. Fully blocked configurations form isolated vertices in $\mathcal{G}$. 
\section{An exactly solvable ordered phase}
\label{sec:FrozenPhase}
For $\pi/4<\theta<3\pi/4$, on the active subspace, both single-bond eigenvalues  $\sin\theta\pm\cos\theta$ are strictly positive. The Hamiltonian is therefore a sum of positive-semidefinite bond terms, and its zero-energy ground-state manifold is the intersection of their kernels. Each bond kernel is spanned by Cartesian-basis configurations in which that bond is blocked. Consequently, the zero-energy manifold is spanned by fully blocked configurations, and such configurations also span the ground-state space, when they exist.

\begin{figure*}
	\centering
	\includegraphics[width=\linewidth]{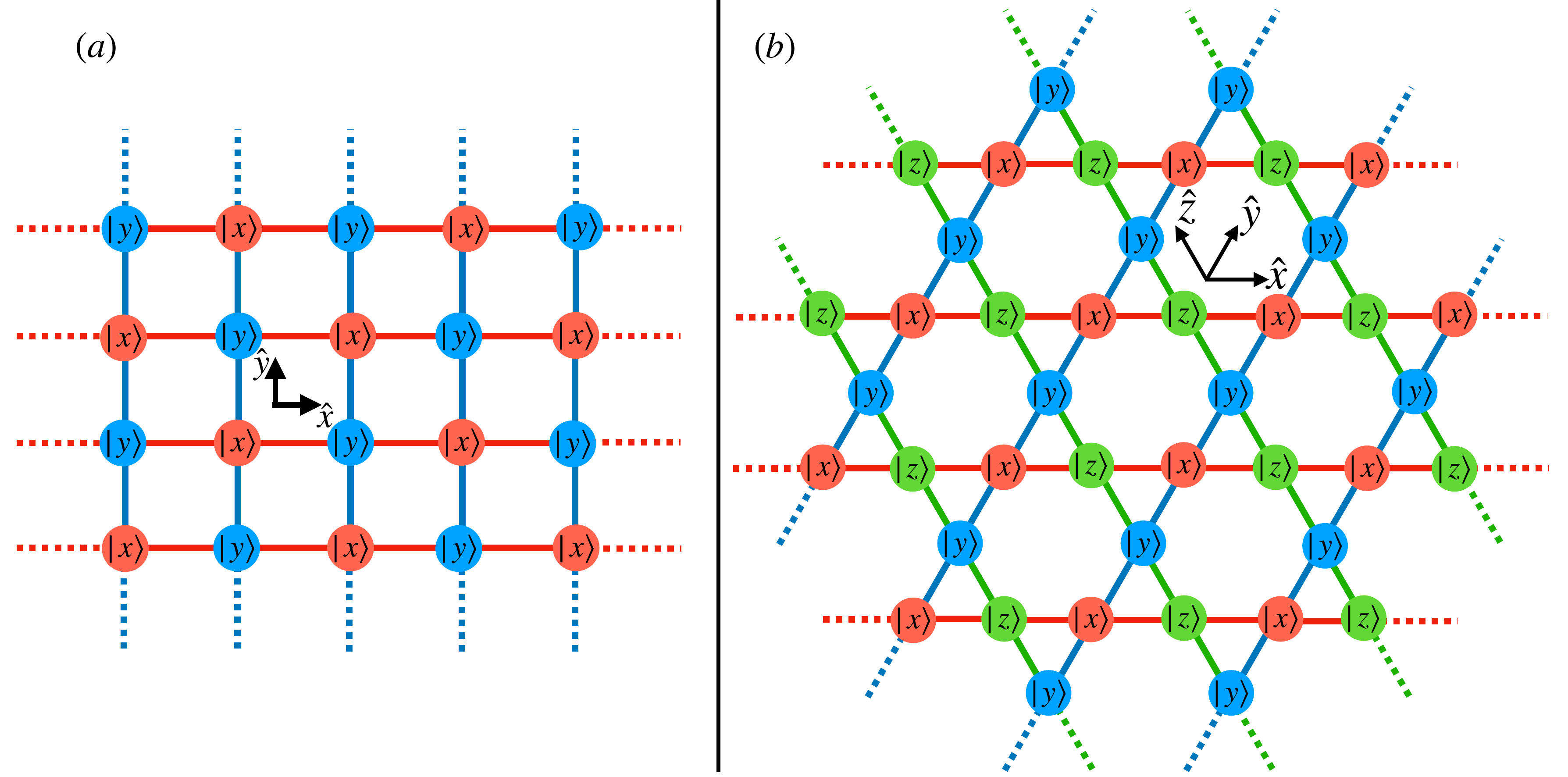}
	\caption{Frozen zero-energy product eigenstates of $H(\theta)$ for any $\theta$: the checkerboard state on the square lattice (left) and the product state on the kagome lattice (right). The checkerboard state has a partner obtained by exchanging $|x\rangle$ and $|y\rangle$; the product state on the kagome lattice has a partner obtained by \textit{one} cyclic permutation of $|x\rangle$, $|z\rangle$, and $|y\rangle$.}
	\label{fig:BlockedStates}
\end{figure*}

On the square lattice, the two fully blocked configurations are the checkerboard arrangement of $\lvert x\rangle$ and $\lvert y\rangle$ states on the $A$ and $B$ sublattices, respectively, and its partner obtained by interchanging $\lvert x\rangle$ and $\lvert y\rangle$. On the kagome lattice, one fully blocked configuration places $\lvert x\rangle$, $\lvert z\rangle$, and $\lvert y\rangle$ on the $A$, $B$, and $C$ sublattices, respectively; its partner follows from a single application of the cyclic permutation $\lvert x\rangle\to\lvert y\rangle\to\lvert z\rangle\to\lvert x\rangle$ (More than one application leads to a state which is not blocked and hence does not lie in the ground state manifold). A schematic of these \textit{frozen} configurations is given in Fig.~\ref{fig:BlockedStates}. Throughout this interval, both models have a two-dimensional ground-state manifold spanned by product states with long-range order. The above statements concerning frozen configurations being the ground state for $\pi/4<\theta<3\pi/4$ does not apply to the square lattice with PBC when either $L_x$ or $L_y$ is odd.
\section{The RK point: $\hat{H}$ as a graph laplacian}
\label{sec:RKPoint}
At $\theta=\pi/4$, the Hamiltonian becomes a sum of bond projectors, $\hat{H}(\pi/4)=\sqrt{2}\sum_b\hat{P}_b$, where $\hat{P}_b=\tfrac{1}{2}[\hat{S}^{\alpha_b}_{i_b}\hat{S}^{\alpha_b}_{j_b}+(\hat{S}^{\alpha_b}_{i_b}\hat{S}^{\alpha_b}_{j_b})^2]$. Its ground-state manifold is therefore the common kernel of these projectors, which also contains the two respective frozen states on each lattice introduced in Sec.~\ref{sec:FrozenPhase}. At this point, however, the lower active-subspace eigenvalue of a bond $\sin\theta-\cos\theta$ also vanishes, enlarging each bond's kernel and hence allowing for additional states in the ground-state manifold beyond the frozen configurations.

To make it convenient to identify the enlarged ground state manifold we perform a similarity transformation on the Hamiltonian in Eq.~\eqref{eq:CompassHamiltonian}. The unitary operator is defined as,
\begin{equation}
	\hat{U}=\bigotimes_i \hat u_i,
	\qquad
	\hat u_i=\mathrm{diag}\bigl(g_i(x),g_i(y),g_i(z)\bigr),
	\label{eq:Combinedgauge}
\end{equation}
diagonal in the Cartesian basis with all entries having unit-modulus. It is a product of on-site unitaries. The choice of $\hat U$ will differ between the two lattices and for each lattice it will further depend on the underlying sublattice as well. On the square lattice, we take $g_i(x)=g_i(y)=1$ on both $A$ and $B$
sublattices, with $g_i(z)=+i$ for $i\in A$ and $g_i(z)=-i$
for $i\in B$. On the kagome lattice the three sublattices $A$, $B$ and $C$ carry phases of:
\begin{equation}
	\begin{aligned}
		(g_A(x),g_A(y),g_A(z))&=(1,1,i),\notag\\
		(g_B(x),g_B(y),g_B(z))&=(-i,1,-i),\notag\\
		(g_C(x),g_C(y),g_C(z))&=(1,-i,-i).
	\end{aligned}
	\label{eq:KagomePhases}
\end{equation}
 The sublattice convention on both the lattices for the corresponding directions of chosen bond couplings is shown in Fig.~\ref{fig:CompassKitaevSchematic}. On performing the similarity transformation on $\hat{H}$ we get,
\begin{equation}
	\langle s|{\hat{U}^{\dagger}\hat{H}\hat{U}}|s'\rangle
	=({g(s)})^*\,g(s')\langle s|{\hat{H}}|s'\rangle,
	\label{eq:SimilaritTrans}
\end{equation}

with $g(s)=\prod_i g_i(s_i)$. It is shown in the appendix that using this similarity transformation with $\hat{U}$ composed of on-site unitaries specified earlier for the two lattices respectively, reverses the sign of every
exchange amplitude and preserves that of every pairing amplitude of Eq.~\eqref{eq:moves}, so that
\emph{both} classes of moves due to the bilinear terms present in Eq.~\eqref{eq:CompassHamiltonian} (in the active subspace)  acquire amplitude $-\cos(\theta)$. Combined with the
fact that each active bond contributes one move in the configuration-graph of Cartesian-basis product vertices (due to the bilinear term) and one unit of diagonal amplitude equal to $\sin(\theta)$ (due to the biquadratic term),  gives the central identity of this work,
\begin{equation}
	\hat{U}^{\dagger}\,\hat{H}(\theta)\,\hat{U}
	=\,\widetilde{H}(\mathcal{G},\theta),
	\quad
	\widetilde{H}(\mathcal{G},\theta)=\sin(\theta)D(\mathcal{G})-\cos(\theta)A(\mathcal{G}),
	\label{eq:Laplacian}
\end{equation}
where $\widetilde{H}(\mathcal{G},\theta)$ is the adjacency-degree representation of the transformed Hamiltonian on the configuration graph
$\mathcal{G}$ with $3^N$ vertices. $A(\mathcal{G})$ is its adjacency matrix and $D(\mathcal{G})=\mathrm{diag}[D(s)]$ is its degree matrix.

At $\theta=\pi/4\equiv \theta_c$, and by unitary equivalence at $\theta=3\pi/4$, the models realize an exact RK point, with $\widetilde{H}(\mathcal{G},\pi/4)=\mathcal{L}(\mathcal{G})/\sqrt{2}$, the graph Laplacian on $\mathcal{G}$~\cite{Mohar1991,GodsilRoyleLaplacian2001}. Thus, $-\mathcal{L}(\mathcal{G})$ generates a continous time random walk on the Cartesian-basis configurations in the transformed basis. The graph Laplacian at $\theta_c$ is positive semidefinite, and its kernel contains one equal-amplitude state per connected component of $\mathcal{G}$, which also span the ground state manifold~\cite{Mohar1991,GodsilRoyleLaplacian2001}. In the original basis, these ground states acquire the configuration-dependent phases specified by $\hat{U}$. Since every line-charge sector is non-empty and distinct sectors are dynamically disconnected, this implies at least $2^{(O(L))}$ connected components in $\mathcal{G}$ on both the lattices and hence the ground state degeneracy is at least $2^{(O(L))}$ in both cases. The frozen configurations are isolated vertices in $\mathcal{G}$.

On either lattice, a ground state in a particular line-charge sector can be constructed by projecting the product state $\ket{P}=\hat{U}\bigotimes_i[(\ket{x_i}+\ket{y_i}+\ket{z_i})/\sqrt{3}]$ onto the corresponding line-charge eigenvalues ($\hat{U}$ is different for the two lattices, though both are composed of on-site unitaries, diagonal in the Cartesian basis). For line charge eigenvalues $\vec{q}=\{q_\ell\}$, this gives $\ket{\Psi^{\vec{q}}}\propto\prod_\ell\tfrac12(1+q_\ell\hat{Q}_\ell)\ket{P}$, where $\ell$ runs over all conserved $\mathbb{Z}_2$ line charges. When the sector consists of a single connected component, this projection yields its unique ground state. In sectors containing multiple such components, it yields a superposition of their ground states. These sector-projected states admit a PEPS
representation~\cite{VerstraeteCirac2004PEPS,
	VerstraeteMurgCirac2008,Cirac2021MPSPEPS}.
Each site carries a tensor with one physical index of dimension
three, corresponding to the three states, and four
virtual indices of dimension two. The virtual indices form two
pairs, one for each of the two charge lines intersecting at the
site, and carry the binary variables introduced by expanding
the corresponding $\mathbb{Z}_2$ charge projectors. Distinct sublattices have distinct tensors.
%
A detailed discussion of the PEPS representation on both the lattices is given in the appendix.

Let $n^\mu_\lambda(s)$ denote the number of sites on sublattice $\mu$ carrying state ${\lambda}$ in configuration $\ket{s}$, and let $n_z(s)$ be the total number of $z$ states present in $\ket{s}$. The normalized ground state associated with a connected component $\mathcal{F}$ of $\mathcal{G}$ is
\begin{align}
	\ket{\Psi_{S,\mathcal{F}}}
	&=\frac{1}{\sqrt{|\mathcal{F}|}}\sum_{s\in\mathcal{F}}
	i^{\,n_z(s)}(-1)^{\,n^B_z(s)}\ket{s},
	\label{eq:PsiSquare}\\[2pt]
	\ket{\Psi_{K,\mathcal{F}}}
	&=\frac{1}{\sqrt{|\mathcal{F}|}}\sum_{s\in\mathcal{F}}
	i^{\,n_z+n^B_x+n^C_y}
	(-1)^{\,n^B_x+n^B_z+n^C_y+n^C_z}\ket{s},
	\label{eq:PsiKagome}
\end{align}
on the square and kagome lattices, respectively. The $(s)$ dependence of the exponents is suppressed in Eq.~\eqref{eq:PsiKagome}. On both lattices the configuration amplitudes are pure phases, admitting a formal representation as square-root Boltzmann factors at imaginary inverse temperature. A similar construction in Ref.~\cite{Changlani_2018} projects
three-color product ground states of the spin-$1/2$ kagome
antiferromagnet onto fixed total-$S^z$ sectors. Though in that case these states can
overlap, if they are in the same $S^z$ sector, and become linearly dependent, and their span need not
exhaust the ground-state manifold. Here, the configuration-graph Laplacian mapping in the orthonormal
Cartesian basis establishes exactly one ground state per connected
component, with states from distinct components mutually orthogonal.
Determining the full ground-state manifold therefore reduces
to characterizing the connected components of the configuration-graph.
	
The sublattice-dependent unitary transformation used for the square lattice also applies to the analogous $S=1$ Kitaev chain with bilinear and biquadratic couplings~\cite{Raja_Ganesh_2026,Sen_etal_PRB_2010,SharmaPandeyInPreparation}, whose sites alternate between the $A$ and $B$ sublattices. In this case, the conserved quantities are bond charges rather than line charges, taking the form $e^{i\pi\hat{S}^{\beta}_{i_b}}e^{i\pi\hat{S}^{\beta}_{i_{b+1}}}$ on a bond $b$ of type $\alpha$. A length $L$ periodic chain has $L$ conserved $\mathbb{Z}_2$ bond charges, defining $2^L$ charge sectors, with ground-state wavefunctions of the same form as those for the square lattice, given in Eq.~\eqref{eq:PsiSquare}. Just as the square- and kagome-lattice ground states admit PEPS representations with bond dimension two, the chain ground states admit an MPS representation with bond dimension two, presented in the Appendix. Using spin fractionalization, Ref.~\cite{Raja_Ganesh_2026} obtained related ground-state wavefunctions for the $S=1$ Kitaev chain. 

Finally, using a variational bound, we can show that the RK endpoints host first-order phase transitions. In the product state $\ket{P}$, each bond is active with probability $4/9$, giving the biquadratic bond operator of $\hat{H}$ in Eq.~\eqref{eq:CompassHamiltonian} an expectation value of $4/9$. Since $\hat{H}(\pi/4)\ket{P}=0$, the corresponding bilinear bond operator's expectation value is $-4/9$. Thus,
\begin{equation}
	\begin{aligned}
		E_0(\theta)
		&\leq \bra{P}\hat{H}(\theta)\ket{P}\\
		&=\frac{4}{9}N_b\bigl(\sin\theta-\cos\theta\bigr)<0,
	\end{aligned}
	\label{eq:bound}
\end{equation}
for $0\leq\theta<\pi/4$, where $N_b$ is the total number of bonds. By contrast, $E_0(\theta)=0$ throughout $\pi/4\leq\theta\leq3\pi/4$. The bound forces a nonzero left-hand slope at $\theta_c=\pi/4$ of the energy density $e_0= E_0/N_b$, whereas its right-hand slope vanishes, establishing a first-order transition. Unitary equivalence implies an analogous transition at $\theta=3\pi/4$. At either endpoint, the ground-state degeneracy is at least exponential in the linear system size and collapses to two upon entering the intervening frozen phase with long range order.

\section{Exact Diagonalization}
\label{sec:ED}
\begin{figure}
	\centering
	\includegraphics[width=\linewidth]{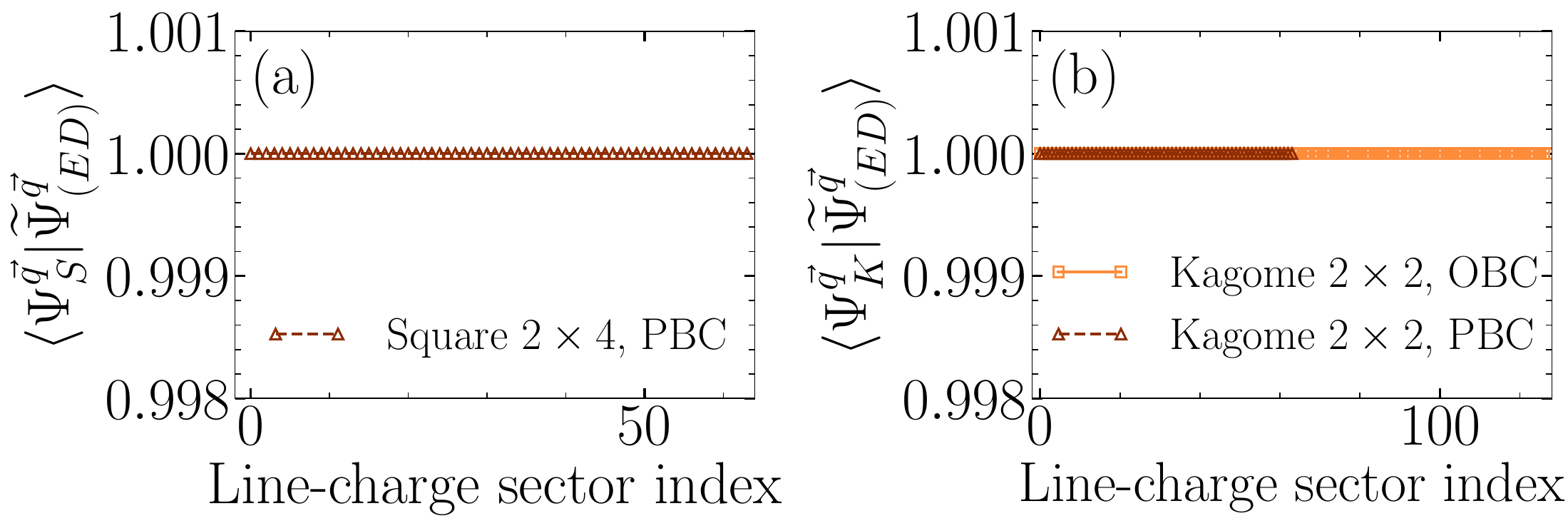}
	\caption{Overlap between the numerical zero-energy state
		$|\widetilde{\Psi}_{\rm ED}^{\vec q}\rangle$ and the analytic
		wavefunction in the same line-charge sector $\vec q$
		[Eqs.~\eqref{eq:PsiSquare} and \eqref{eq:PsiKagome}]
		for a $2\times4$ square lattice with PBC (left) and
		$2\times2$ kagome lattice with OBC and PBC (right).}
	\label{fig:SquareKagomeOverlap}
\end{figure}

To test the analytic ground-state wavefunctions in
Eqs.~\eqref{eq:PsiSquare} and \eqref{eq:PsiKagome} and the
predicted first-order transition at $\theta=\pi/4$, we perform
exact diagonalization (ED) of $\hat{H}$ in  Eq.~\eqref{eq:CompassHamiltonian}
on a $2\times4$ square lattice with PBC and $2\times2$ kagome
lattices with OBC and PBC.
For these clusters, each line-charge sector contains one
non-frozen connected component; though the two frozen configurations
form additional isolated components within a sector. Fig.~\ref{fig:SquareKagomeOverlap} shows the overlap of the predicted ground-states in Eq.~\eqref{eq:PsiSquare} and Eq.~\eqref{eq:PsiKagome} with the ground state obtained by ED for all the sectors. In the sectors containing the frozen states, the frozen states are projected out of the
numerical zero-energy subspace obtained by ED before comparison with the
analytic state supported on the non-frozen component of that sector.
Fig.~\ref{fig:SquareKagomeOverlap} shows that the overlap
is unity in every sector for all three cases.

\begin{figure}
	\centering
	\includegraphics[width=\linewidth]{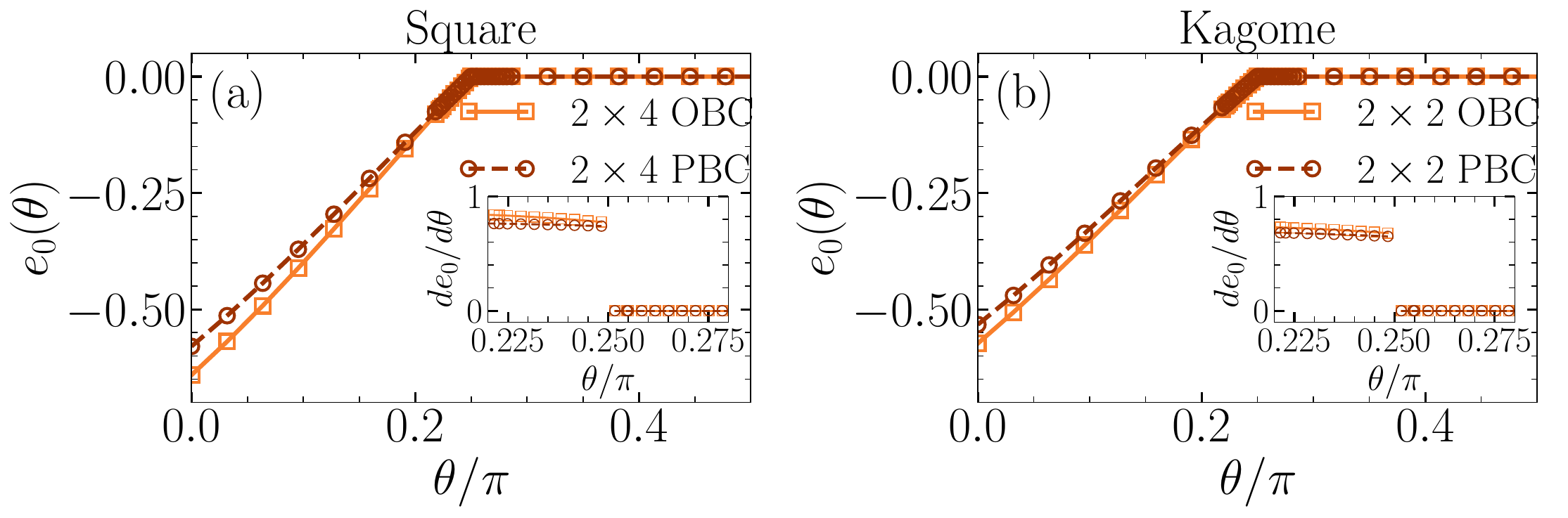}
	\caption{Ground-state energy density $e_0$ of the Hamiltonian in Eq.~\eqref{eq:CompassHamiltonian} as a function of $\theta/\pi$ for a $2\times4$ square lattice (left) and a $2\times2$ kagome lattice (right), with both OBC and PBC. Insets show $de_0/d\theta$; its discontinuity at $\theta/\pi=0.25$ signals a first-order transition.}
	\label{fig:SquareKagomeEnergyWithDerivative}
\end{figure}

Fig.~\ref{fig:SquareKagomeEnergyWithDerivative} shows the ground-state energy density $e_0\equiv E_0/N_b$ (where $E_0$ is the ground state energy and $N_b$ is the number of bonds present in the lattice) as a function of $\theta/\pi$ for the square and kagome lattices (here we also include square-lattice OBC data). For $\theta<\pi/4$, $e_0<0$ and the ground state is nondegenerate. For $\pi/4<\theta<3\pi/4$, $e_0=0$ and the ground-state manifold is spanned by two frozen states on each lattice mentioned in Sec.~\ref{sec:FrozenPhase}, and shown in the schematic in Fig.~\ref{fig:BlockedStates}. The insets show a discontinuity in $de_0/d\theta$ at $\theta=\pi/4$, consistent with the first-order transition predicted in Sec.~\ref{sec:RKPoint}.

\section{Discussion}
Our results establish a connection between subsystem-symmetry induced Hilbert-space decomposition~\cite{Budde_preprint_2026} and RK physics~\cite{RokhsarKivelson1988,Henley_2004,Castelnovo_2005} in $S=1$ compass models on square and kagome lattices~\cite{NussinovVandenBrink2015,Nussinov_Fradkin_PRB_2005,Dorier_etal_PRB_2005,Yang_etal_PRB_2020}, having both bilinear and biquadratic interactions, with their relative strengths parametrized by an angle $\theta$. Using a transformation built of on-site unitary operators, the Hamiltonian in Eq.~\eqref{eq:CompassHamiltonian} on both the lattices takes the adjacency-degree representation given in Eq.~\eqref{eq:Laplacian} on a configuration graph $\mathcal{G}$, whose vertices represent product  states in the $S=1$ Cartesian basis. In both cases the ground state manifold is known exactly in an extended region of the parameter space. At the two exactly solvable endpoints of this interval, the Hamiltonian transforms into the Laplacian of $\mathcal{G}$ in the Cartesian basis~\cite{Mohar1991,GodsilRoyleLaplacian2001,Sen_etal_PRB_2010,SharmaPandeyInPreparation}. Ground-state counting and corresponding wavefunction construction thereby reduce to a graph-connectivity problem: each connected component of the configuration graph supports one state in the ground-state manifold, an equal-amplitude superposition over the states in the connected component in the transformed basis, with additional configuration-dependent phases in the original basis~\cite{Mohar1991,GodsilRoyleLaplacian2001}. Both lattices possess $m=O(L)$ conserved $\mathbb{Z}_2$ line charges, yielding $2^{m}$ symmetry sectors. This gives a lower bound on the number of connected components in the configuration-graph and hence on the ground state degeneracy at the RK points. The states in the ground state have a PEPS representation of bond dimension two on both the lattices.

Between the two RK endpoints, the ground-state manifold is spanned
by two frozen configurations, yielding an ordered phase on both
lattices. The discontinuity in the first derivative of the
ground-state energy density establishes a first-order transition
at each endpoint. The RK points thus lie at the boundaries of an ordered phase, reminiscent of the role of RK points in QDMs~\cite{MoessnerSondhi2001,Misguich2002,Syljuasen2006,MoessnerRaman2011}.

The adjacency-degree representation of the Hamiltonian in Eq.~\eqref{eq:Laplacian} also makes the analogy with QDMs explicit~\cite{RokhsarKivelson1988,MoessnerSondhi2001,Misguich2002,Syljuasen2006,MoessnerRaman2011}. Active bonds play the role of flippable plaquettes: in the transformed Cartesian basis, the bilinear interaction generates transitions between configurations, while the biquadratic interaction assigns a diagonal energy to each active bond. These therefore correspond to the kinetic and potential-energy terms of a QDM, respectively. In two-dimensional
QDMs on a $L\times L$ torus, integer winding numbers label $O(L^2)$
sectors on bipartite lattices, whereas winding parities label finitely
many sectors on non-bipartite lattices. By contrast, the compass
models considered here have $m=O(L)$ independent and mutually commuting conserved
$\mathbb{Z}_2$ line charges labeling $2^m$ sectors.
In either setting, additional fragments within a symmetry sector, including frozen configurations, can occur. This analogy motivates studying additional spin interactions that modify the ground-state configuration ensemble or its amplitudes, and investigating whether the resulting models can support Coulomb correlations and an emergent gauge description. Determining whether such behavior is confined to fine-tuned points or can persist throughout an extended phase remains an open question.

At the RK endpoints, our construction realizes the equal-weight case of the stochastic-matrix-form framework of Ref.~\cite{Castelnovo_2005}, applied separately to each connected fragment. The transformed Hamiltonian is a graph Laplacian, so its negative directly generates a classical Markov process with a uniform stationary distribution within each fragment. Within each connected component, its relaxation rates coincide with the
quantum excitation energies~\cite{Henley_2004,Castelnovo_2005}. Thus, the microscopic spin Hamiltonian acquires a direct stochastic interpretation through an on-site phase transformation alone, without any additional reweighting of configuration amplitudes. Finding exact excited states is another direction to be explored in the future. 

The Laplacian representation at the RK endpoints also identifies
continuous families of local Hamiltonians with exactly the same
ground-state manifold. Specifically, replacing the uniform bond
couplings by arbitrary strictly positive coefficients gives
\begin{equation}
	\hat H_{\sigma}(\{J_b\})
	=
	\sum_b J_b
	\left[
	\left(
	\hat S_{i_b}^{\alpha_b}
	\hat S_{j_b}^{\alpha_b}
	\right)^2
	+
	\sigma\,
	\hat S_{i_b}^{\alpha_b}
	\hat S_{j_b}^{\alpha_b}
	\right],
	\qquad J_b>0,
	\label{eq:weighted_RK_family}
\end{equation}
where $\sigma=+1$ and $\sigma=-1$ preserve the ground-state
manifolds at $\theta=\pi/4$ and $3\pi/4$, respectively.
This freedom includes directional anisotropy and arbitrary
positive bond disorder as long as at each bond, the biquadratic coupling equals the
bilinear coupling at $\theta=\pi/4$ and its negative at $\theta=3\pi/4$. Each Hamiltonian remains a positive sum of the same bond
projectors, so their common kernel is unchanged. Equivalently,
after the appropriate on-site unitary transformation, these
modifications change the edge weights of the configuration-graph
Laplacian while retaining its connected components and the uniform
stationary distribution within each component. Further local extensions can make the transition strengths depend
on nearby state occupations, provided the forward and reverse
rates remain equal and strictly positive. Such terms introduce
interactions involving additional sites while retaining the
original two-spin moves and their configuration-dependent
ground-state phases. Allowing rates to vanish can modify the ground state degeneracy as this could modify the number of connected components of the configuration graph.

The Cartesian-basis construction is independent of spatial dimension
and works at fixed $S=1$ without imposing a relation between spin and
coordination number. The present construction requires compatible
on-site phases that transform the Hamiltonian into an adjacency-degree representation, given in Eq.~\eqref{eq:Laplacian}, on a 
configuration-graph. As long as at most two types of couplings meet at a site, the given construction holds, provided the subgraphs of every coupling are bipartite (for example the $S=1$ Kitaev chain, in fact exactly the same choice of $\hat{U}$ used for the square lattice compass model works for the chain as well). In particular this  construction to identify an appropriate $\hat{U}$ fails when all three bond types meet at a site, as in the $S=1$
honeycomb Kitaev model, because the local phase constraints are
inconsistent. In that case a more complicated $\hat{U}$ might exist, finding which will be a direction for the future.

Characterizing the phase of the $S=1$ compass models on both the lattices
outside the extended solvable interval remains another open direction for future work. In particular at $\theta=0$ Eq.~\eqref{eq:Laplacian} gives
$\hat U^\dagger \hat H(0)\hat U=-A$, where $A$ is the
configuration-graph adjacency matrix. Thus, the question of finding the ground state energy and wavefunction at $\theta = 0$ might be inferred from studying the structure of $\mathcal{G}$ which will also be focus of future work.  

More broadly, these models illustrate how RK physics can arise directly in microscopic spin Hamiltonians with only nearest-neighbour two-site interactions. A product of on-site unitaries exposes the exact adjacency--degree representation in Eq.~\eqref{eq:Laplacian} for arbitrary $\theta$, with the RK points emerging as special cases where the Hamiltonian is unitarily equivalent to a graph Laplacian. Identifying other spin models that admit such local transformations offers a route to further exact RK constructions and a possible graph-theoretic approach to their properties away from the solvable points.
\begin{acknowledgments}
This work began after the author read the arXiv preprint of Ref.~\cite{Raja_Ganesh_2026} and explored possible extensions, including through discussions with Claude Opus 4.8. In particular, Claude Opus 4.8 provided insights into the properties of graph Laplacians and their implications for the ground-state manifold. Claude Opus 4.8 and ChatGPT Sol 5.6 assisted in writing the exact diagonalization code. These models, together with ChatGPT Sol 6.1, were also used at various stages of the project to assist with literature searches and polishing the language of the manuscript.  The author thanks Ayush Sharma for collaboration on related work~\cite{SharmaPandeyInPreparation}, which also motivated the present study; Kedar Damle for his guidance and encouragement; and Sanjay Moudgalya for useful comments on an earlier draft. This work was supported by the Department of Atomic Energy, Government of India, under Project Identification No.~RTI-4012. The author takes full responsibility for all the results and claims.
\end{acknowledgments}
\FloatBarrier

\bibliography{bibliography}
\onecolumngrid
\bigskip
\begin{center}\rule{0.45\textwidth}{0.4pt}\end{center}
\medskip
\twocolumngrid

\setcounter{equation}{0}
\renewcommand{\theequation}{AP\arabic{equation}}
\renewcommand{\theHequation}{AP\arabic{equation}} 

\section*{Appendix}

\subsection{Construction of the unitary transformation}
We derive the expression of $\hat{U}$ in Eq.~\eqref{eq:Combinedgauge} for both square and kagome lattices at once. Let
$(\alpha,\beta,\gamma)$ be cyclic and consider a bond of type $\alpha$ joining sites
$i$ and $j$. We will work in the $S=1$ Cartesian basis. For a product state $\bigotimes_k\ket{\phi_k}$ with
$\ket{\phi_k}=a_k\ket{x}+b_k\ket{y}+c_k\ket{z}$, write $u_k=\braket{\beta}{\phi_k}$
and $v_k=\braket{\gamma}{\phi_k}$ ($k$ runs over all the lattice sites. $a_k$, $b_k$ and $c_k$ are all non-zero and have unit modulus). 

To make sure that $\hat{H}$ takes the adjacency-degree form over the configuration graph given in Eq.~\eqref{eq:Laplacian}, we want to make sure that the amplitude of the two types of moves due to the bilinear terms of $\hat{H}$ on every bond on both the lattices given in Eq.~\eqref{eq:moves} are exactly the same. Thus, we want both the pairing move and the exchange move to have the same sign. To enforce this we choose the transformation $\hat{U}$ such that, after the transformation, the exchange terms get an extra minus sign while the pairing terms are left invariant. Hence, we require antisymmetry under the exchange moves and equal amplitudes under the pairing moves. Using $v_k$ and $u_k$ in Eq.~\eqref{eq:moves} we find,
\begin{equation}
	u_iv_j=-v_iu_j,
	\qquad
	u_iu_j=v_iv_j .
	\label{eq:EMlocal}
\end{equation}
With the ratio $r^{\alpha}_k\equiv v_k/u_k$ these equations become $r^{\alpha}_j=-r^{\alpha}_i$
and $r^{\alpha}_ir^{\alpha}_j=1$, whence
\begin{equation}
	r^{\alpha}_i=\pm i,\qquad r^{\alpha}_j=-r^{\alpha}_i .
	\label{eq:EMratio}
\end{equation}
Thus, $\hat{H}$ takes the adjacency-degree representation on the configuration graph in the basis of product Cartesian states, if on every
bond of type $\alpha$, the associated ratios $r_i^{\alpha}$ and $r_j^{\alpha}$ equal $\pm i$ and take opposite values at the two
endpoints of $\alpha$. The three ratios at a site, $r^{x}_i=c_i/b_i$, $r^{y}_i=a_i/c_i$ and
$r^{z}_i=b_i/a_i$, are not independent but obey
\begin{equation}
	r^{x}_ir^{y}_ir^{z}_i=1 .
	\label{eq:EMsite}
\end{equation}

On the square lattice only $x$- and $y$-type bonds occur and meet at a site, so only $r^x$ and
$r^y$ are constrained. Taking $a_{(m,n)}=b_{(m,n)}=1$ and
$c_{(m,n)}=i(-1)^{m+n}$ gives $r^{x}={(r^{y})}^{-1}=i(-1)^{m+n}$, which reverses
sign under $m\to m+1$ and under $n\to n+1$ alike, satisfying
Eq.~\eqref{eq:EMratio} on every bond (here we have used two indices $(m,n)$ to specify the coordinate of a site to make it convenient to specify the sublattice dependent phase). The result depends on the site only
through the parity of $m+n$, giving the two-sublattice dependent phases quoted for the square lattice below Eq.~\eqref{eq:Combinedgauge} in Sec.~\ref{sec:RKPoint}.

On the kagome lattice the $A$--$B$, $B$--$C$ and $C$--$A$ bonds are of type $x$,
$z$ and $y$, so each sublattice is touched by only two out of the three bond types:
$A$ by $x$ and $y$, $B$ by $x$ and $z$, $C$ by $z$ and $y$. Since every bond
connects distinct sublattices, the alternation required by
Eq.~\eqref{eq:EMratio} is realized by a uniform assignment on each sublattice. Choosing $r^{x}_A=r^{z}_B=r^{y}_C=i$ fixes
$r^{x}_B=r^{z}_C=r^{y}_A=-i$, and Eq.~\eqref{eq:EMsite} then forces the
remaining, unconstrained ratio at each site to be,
$r^{z}_A=r^{y}_B=r^{x}_C=1$. Fixing the overall phase of each
single-site state yields Eq.~\eqref{eq:KagomePhases}.

\subsection{Verification of the Hamiltonian's adjacency-degree representation}
\label{app:LaplacianIdentity}
Write $g(s)=\prod_ig_i(s_i)$ in Eq.~\eqref{eq:SimilaritTrans}, where $i$ runs over all the sites of the lattice. For an exchange move due to a bilinear term of $\hat{H}$, on a bond of type $\alpha$
joining sites $i$ and $j$, the two configurations $|s\rangle$ and $|s'\rangle$ differ by
$(s_i,s_j)=(\beta,\gamma)\to(\gamma,\beta)$, so
$({g(s)})^*g(s')=({g_i(\beta)g_j(\gamma)})^*\,g_i(\gamma)g_j(\beta)
={r^{\alpha}_i/r^{\alpha}_j}=-1$ by Eq.~\eqref{eq:EMratio}. For a
pairing move, $(\beta,\beta)\to(\gamma,\gamma)$ and
$({g(s)})^*g(s')=({r^{\alpha}_ir^{\alpha}_j})=+1$, due to Eq.~\eqref{eq:EMratio}. Since the bare contributions are $+\cos\theta$ and $-\cos\theta$ respectively
[Eq.~\eqref{eq:moves}], both transformed off-diagonal elements due to the bilinear terms of $\hat{H}$ equal
$-\cos\theta$. The diagonal element due to the biquadratic term of $\hat{H}$ is $\sin\theta\,D(s)$, where $D(s)$ is the number of active bonds in $|s\rangle$
and each active bond contributes exactly one off-diagonal partner, so the degree
of $s$ in $\mathcal{G}$ is also $D(s)$ and Eq.~\eqref{eq:Laplacian} follows.

\subsection{Bond dimension two PEPS}
\label{sec:PEPS}
In this appendix we give a PEPS representation of the ground-state wavefunctions of $\hat{H}$ at the RK point. At that point a ground state in a line-charge sector
$\vec{q}$ on either the kagome or the square lattice (or a superposition of states in the ground state manifold when the sector contains more than one connected component) is obtained
by projecting $\ket{P}$ onto that sector:
\begin{equation}
	\ket{\Psi^{\vec{q}}}
	\propto
	\prod_{\ell=1}^{m}
	\frac{1+q_\ell\hat{Q}_\ell}{2}\ket{P}.
	\label{eq:ProjectedWavefn}
\end{equation}
Here $\ell$ runs over the conserved $\mathbb{Z}_2$ line charges
$\hat{Q}_\ell$, $q_\ell\in\{-1,1\}$ specifies the desired
eigenvalue of each charge, and
\begin{equation}
	\ket{P}
	=
	\hat{U}\bigotimes_i
	\frac{\ket{x_i}+\ket{y_i}+\ket{z_i}}{\sqrt{3}}
	=
	\frac{1}{3^{N/2}}\sum_s g(s)\ket{s}.
\end{equation}
The sum runs over all $3^N$ Cartesian-basis product configurations
$s=(s_1,\ldots,s_N)$, and $g(s)=\prod_i g_i(s_i)$ contains
the on-site phases specified in Sec.~\ref{sec:RKPoint} of the main-text for the two lattices.

For a fixed configuration $s$, the numerator of
each line-charge projector can be expanded as
\begin{equation}
	1+q_\ell Q_\ell(s)
	=
	\sum_{t_\ell=0}^{1}
	q_\ell^{t_\ell}
	\prod_{i\in\ell}
	\bigl[w_{\tau_{\ell i}}(s_i)\bigr]^{t_\ell}.
	\label{eq:tn_projector_expansion}
\end{equation}
Here $Q_\ell(s)$ is the eigenvalue of $\hat{Q}_\ell$ on
$\ket{s}$, $\tau_{\ell i}$ denotes the rotation component
used by line $\ell$ at site $i$, and we have used
\begin{equation}
	e^{i\pi\hat{S}^{\alpha}}\ket{\beta}
	=
	(2\delta_{\alpha\beta}-1)\ket{\beta}
	=
	w_\alpha(\beta)\ket{\beta}.
\end{equation}

Each site of either lattice lies at the intersection of two
$\mathbb{Z}_2$ charge lines. We introduce two virtual indices
$a_j,b_j\in\{0,1\}$ along each incident line $j=1,2$.
Writing $\tau_1$ and $\tau_2$ for the rotation components
of these two lines at site $i$, we define the local tensor
\begin{equation}
	T_i^{s_i}{}_{a_1b_1a_2b_2}
	=
	g_i(s_i)\,
	\delta_{a_1b_1}\delta_{a_2b_2}
	w_{\tau_1}(s_i)^{a_1}
	w_{\tau_2}(s_i)^{a_2}.
	\label{eq:tn_local_peps}
\end{equation}
The tensor has one physical index of dimension three and
four virtual indices, each of dimension two. Contracting
neighboring virtual indices along the lattice bonds,
together with the Kronecker deltas in
Eq.~\eqref{eq:tn_local_peps}, propagates one common binary
variable along each line.

For a closed line $\ell$, the desired charge $q_\ell$ is
implemented by inserting
\begin{equation}
	B(q_\ell)
	=
	\begin{pmatrix}
		1&0\\
		0&q_\ell
	\end{pmatrix}
	\label{eq:tn_charge_insertion}
\end{equation}
on any one virtual bond of that line. For an open line,
the same factor is obtained by contracting the virtual
indices at its two ends with $(1,1)$ and $(1,q_\ell)$,
respectively.

The resulting tensor contraction for a fixed physical configuration $s$ is
\begin{align}
	\operatorname{Contr}\{T,B\}(s)
	&=
	g(s)
	\prod_{\ell=1}^{m}
	[1+q_\ell Q_\ell(s)]
	\notag\\
	&=
	2^m g(s)\,
	\mathbf{1}_{\Omega_{\vec q}}(s).
	\label{eq:tn_peps_amplitude}
\end{align}
Here $\Omega_{\vec q}$ is the set of Cartesian-basis product
configurations satisfying $Q_\ell(s)=q_\ell$ for every
line $\ell$. The indicator
function $\mathbf{1}_{\Omega_{\vec q}}(s)$ equals one
when $s\in\Omega_{\vec q}$ and zero otherwise.

If $\mathcal N_{\vec q}=|\Omega_{\vec q}|$
is the number of configurations in $\Omega_{\vec q}$ then multiplying the contraction in
Eq.~\eqref{eq:tn_peps_amplitude} by
$(2^m\sqrt{\mathcal N_{\vec q}})^{-1}$ gives
the normalized wavefunction amplitude of $\ket{s}$.
Summing these amplitudes over all configurations
reproduces the normalized projected state in
Eq.~\eqref{eq:ProjectedWavefn}:
\begin{equation}
	\ket{\Psi^{\vec q}}
	=
	\frac{1}{2^m\sqrt{\mathcal N_{\vec q}}}
	\sum_s
	\operatorname{Contr}\{T,B\}(s)\ket{s}
	=
	\frac{1}{\sqrt{\mathcal N_{\vec q}}}
	\sum_{s\in\Omega_{\vec q}}g(s)\ket{s}.
\end{equation}

For the square lattice, take horizontal bonds of type $x$
and vertical bonds of type $y$. The row and column charges
use rotations $e^{i\pi\hat{S}^y}$ and
$e^{i\pi\hat{S}^x}$, respectively. With sites labelled
by integers $(m,n)$, a compatible phase choice is
\begin{equation}
	(g_i(x),g_i(y),g_i(z))
	=
	\bigl(1,1,i(-1)^{m+n}\bigr).
	\label{eq:tn_square_phases}
\end{equation}
The local tensor is therefore
\begin{equation}
	T_i^{s_i}{}_{lrud}
	=
	g_i(s_i)\delta_{lr}\delta_{ud}
	w_y(s_i)^l w_x(s_i)^u,
	\label{eq:tn_square_tensor}
\end{equation}
where $l,r,u,d\in\{0,1\}$ label the left, right, upper,
and lower virtual legs. 

For the kagome lattice, choose the bond convention
$AB:x$, $AC:y$, and $BC:z$. Compatible on-site phases are
\begin{align}
	(g_A(x),g_A(y),g_A(z))&=(1,1,i),\notag\\
	(g_B(x),g_B(y),g_B(z))&=(-i,1,-i),\notag\\
	(g_C(x),g_C(y),g_C(z))&=(1,-i,-i).
	\label{eq:tn_kagome_phases}
\end{align}
Writing $u_\alpha,v_\alpha$ for the two virtual legs
along a line of bond type $\alpha$, the tensors at a
site $i$ on the respective sublattices are
\begin{align}
	T_A^{s_i}{}_{u_xv_xu_yv_y}
	&=
	g_A(s_i)\delta_{u_xv_x}\delta_{u_yv_y}
	w_y(s_i)^{u_x}w_x(s_i)^{u_y},
	\notag\\
	T_B^{s_i}{}_{u_xv_xu_zv_z}
	&=
	g_B(s_i)\delta_{u_xv_x}\delta_{u_zv_z}
	w_z(s_i)^{u_x}w_x(s_i)^{u_z},
	\notag\\
	T_C^{s_i}{}_{u_yv_yu_zv_z}
	&=
	g_C(s_i)\delta_{u_yv_y}\delta_{u_zv_z}
	w_z(s_i)^{u_y}w_y(s_i)^{u_z}.
	\label{eq:tn_kagome_tensors}
\end{align}
Here $A$, $B$, and $C$ denote the three kagome sublattices
shown in Fig.~\ref{fig:CompassKitaevSchematic}(b).

\subsection{MPS for the $S=1$ Kitaev chain}
Consider a periodic even length $S=1$ Kitaev chain with $N$ sites, labelled $i=0,\ldots,N-1$. Bond $i$ connects sites $i$ and $i+1$ and has type $\alpha_i=x$ for even $i$ and $\alpha_i=y$ for odd $i$. We denote the complementary component in $\{x,y\}$ by $\beta_i$. The conserved bond charges are
\begin{equation}
	\hat{Q}_i={e^{i\pi\hat{S}^{\beta_i}}_i}
	{e^{i\pi\hat{S}^{\beta_i}}_{i+1}},
	\quad
	\hat{Q}_i|\Psi^{\vec q}\rangle
	=q_i|\Psi^{\vec q}\rangle,
	\quad q_i\in\{-1,+1\},
	\label{eq:tn_chain_data}
\end{equation}
where $\vec q$ labels the sector specified by the bond-charge eigenvalues $\{q_i\}$. In the Cartesian basis, define $w_\beta(s)$ through
\begin{equation}
	e^{i\pi\hat S^\beta}|s\rangle
	=w_\beta(s)|s\rangle,
	\qquad
	w_\beta(s)=
	\begin{cases}
		+1,&s=\beta,\\
		-1,&s\neq\beta,
	\end{cases}
\end{equation}
The sublattice-dependent on-site phases are
\begin{equation}
	\bigl(g_j(x),g_j(y),g_j(z)\bigr)
	=\bigl(1,1,i(-1)^j\bigr),
	\qquad
	g(s)=\prod_{j=0}^{N-1}g_j(s_j),
\end{equation}
where $s=(s_0,\ldots,s_{N-1})$ denotes a Cartesian-basis configuration.

For each physical index $s_i \in\{x,y,z\}$, introduce a $2\times2$ matrix $A_i^{s_i}$ with virtual indices $u,v\in\{-1,+1\}$:
\begin{equation}
	[A_i^{s_i}]_{uv}
	=
	g_i(s_i)\,
	\delta_{u,w_{\beta_{i-1}}(s_i)}
	\delta_{v,q_i w_{\beta_i}(s_i)}.
	\label{eq:tn_chain_matrices}
\end{equation}
Contracting the right virtual index at site $i$ with the left virtual index at site $i+1$ imposes
$q_i w_{\beta_i}(s_i)=w_{\beta_i}(s_{i+1})$, or equivalently
$w_{\beta_i}(s_i)w_{\beta_i}(s_{i+1})=q_i$. Consequently,
\begin{equation}
	\operatorname{Tr}
	[A_0^{s_0}\cdots A_{N-1}^{s_{N-1}}]
	=
	g(s)
	\prod_{i=0}^{N-1}
	\delta_{q_i,\,
		w_{\beta_i}(s_i)w_{\beta_i}(s_{i+1})}.
	\label{eq:tn_chain_amplitude}
\end{equation}
The amplitude therefore equals $g(s)$ for configurations in the sector $\Omega_{\vec q}$ and vanishes otherwise. For a sector, the normalized state consequently has the exact bond-dimension-two MPS representation
\begin{equation}
	|\Psi^{\vec q}\rangle
	=
	\frac{1}{\sqrt{\mathcal N^{\vec q}}}
	\sum_{s_0,\ldots,s_{N-1}\in\{x,y,z\}}
	\operatorname{Tr}
	[A_0^{s_0}\cdots A_{N-1}^{s_{N-1}}]
	|s_0,\ldots,s_{N-1}\rangle.
	\label{eq:tn_chain_mps}
\end{equation}
\end{document}